\documentclass[runningheads]{llncs}\pdfoutput=1
\usepackage{graphicx}\pdfoutput=1
\usepackage{appendix}\pdfoutput=1
\usepackage{amssymb,amsfonts}\pdfoutput=1
\usepackage{algorithmic}\pdfoutput=1
\usepackage{textcomp}
\usepackage{xcolor}
\usepackage[caption=false]{subfig}
\usepackage{xspace}
\usepackage{enumerate}
\usepackage{subfig}
\usepackage{msc}
\setmsckeyword{} 
\newcommand{\eat}[1]{}

\usepackage{algorithm}
\usepackage{algorithmic}

\newtheorem{protocol}[theorem]{Protocol}

\newcommand{\eps}{\ensuremath{\epsilon}}
\newcommand{\term}[1]{\emph{#1}\xspace}
\newcommand{\eg}{\emph{e.g.},\xspace}
\newcommand{\ie}{\emph{i.e.},\xspace}

\newcommand{\A}{\ensuremath{\mathcal{A}}}
\newcommand{\EBC}[1]{\ensuremath{\mathrm{EBC}(#1)}\xspace}
\newcommand{\PEBC}[1]{\ensuremath{\mathrm{pEBC}(#1)}\xspace}
\newcommand{\Pow}[1]{\ensuremath{\mathcal{P}\left(#1\right)}\xspace}
\newcommand{\Unif}[1]{\ensuremath{\mathrm{Unif}\left(#1\right)}\xspace}
\mathchardef\mhyphen="2D

\newcommand{\mybinom}[2]{%
    \left(
    \begin{array}{@{}c@{\,}} #1\\#2 \end{array}
    \right)}

\newcommand{\ITSampler}{\textsc{InverseTransformSampler}\xspace}
\newcommand{\PFSampler}{\textsc{PickAndFlipSampler}\xspace}

\newcommand{\SubsetRelease}{\textsc{SubsetRelease}\xspace}
\newcommand{\PrivateEBC}{\textsc{PrivateEBC}\xspace}
\newcommand{\ReciprocateandSum}{\textsc{ReciprocateandSum}\xspace}
\newcommand{\PrivateReciprocateandSum}{\textsc{PrivateReciprocateandSum}\xspace}

\newcommand{\PrivatePathCount}{\textsc{PrivatePathCount}\xspace}

\usepackage{xr}
\begin{document}
\title{Assessing Centrality Without Knowing Connections}
\author{Leyla Roohi\orcidID{0000-0002-0589-5814} \and
Benjamin I.P. Rubinstein\orcidID{0000-0002-2947-6980} \and
Vanessa Teague\orcidID{0000-0003-2648-2565}}

\institute{School of Computing and Information Systems, University of Melbourne, Australia \\ 
\email{\{Leyla.Roohi, Benjamin.Rubinstein, Vjteague\}@unimelb.edu.au}
}

\maketitle              

\begin{abstract}
We consider the privacy-preserving computation of
node influence in distributed social networks, as measured by egocentric betweenness centrality (EBC). 
Motivated by modern
communication networks spanning multiple providers, 

we show for the first time how
multiple mutually-distrusting parties can successfully compute node EBC while revealing only differentially-private information
about their internal network connections. 
A theoretical utility analysis upper bounds a primary source of private EBC error---private release of ego networks---with high probability. 
Empirical results demonstrate practical applicability 
with a low 1.07 relative error achievable at strong privacy budget $\epsilon=0.1$ on a Facebook graph, and insignificant performance degradation as the number of network provider parties grows.
\end{abstract}

\section{Introduction}
This paper concerns the measurement of node importance in
communication networks with
\emph{egocentric betweenness centrality} (EBC)~\cite{goh2003betweenness}, representing how much a node's neighbours depend on the node for inter-connection. EBC has emerged as a popular
centrality measure that is used widely in practice~\cite{marsden2002egocentric}.

EBC computation has many  applications.  In conjunction with methods for identifying fake news~\cite{kwon2013prominent,monti2019fake}, EBC can be used
to limit its propagation by targeting interventions at those
individuals who are most critical in spreading information. EBC computation is straightforward when all communication network information is available to one trusted party.
However in reality modern telecommunications involve competing network providers, even within the one country.
 While many people communicate between countries with
completely different networks, where no central authority that can 
view all their connections.   
While recent work~\cite{roohi2019differentiallyprivate} considered the case of two mutually-distrusting networks, 
multiple networks
are  essential for understanding one person's
communication and presents non-trivial technical challenges.

Here we present
a protocol that preserves the privacy of the internal
connections of each of arbitrarily-many networks while they collaborate on the computation of EBC.    
By carefully structuring information flow, we achieve highly accurate results and strong privacy protection.
We produce a private output that can be safely published. 
We assume the complete list of nodes (\ie people) is public, while individual connections are private. 
Each service provider knows the connections within its own network, plus the connections between one of its members and the outside (\eg from when they contact someone in a different network). Connections internal to other networks are unknown. We prove that our protocol preserves \emph{edge differential privacy}~\cite{hay2009accurate} in which the existence or non-existence of an edge
  must be protected.  We present:
\begin{enumerate}
    \item A protocol for multi-party
    private EBC computation;
    \item \label{point2} A strengthened adversarial model in comparison to prior work---all
    
     participating networks are protected
    by edge-DP, even when the final output is published; 
    \item A high-probability utility bound on a core component of our protocol: private distributed release of ego networks; 
    \item Comprehensive empirical validation on open Facebook, Enron and PGP datasets. This demonstrates applicability of our algorithm with modest 1.07 relative error at strong $\epsilon=0.1$ DP, practical runtimes, and insignificant degradation with increasingly many parties. 
    
\end{enumerate}

Near-constant accuracy with increasing numbers of parties is both surprising and significant, as is our innovation over past work~\cite{roohi2019differentiallyprivate} by preventing leakage at final EBC release. 
Our protocol is substantially more efficient than a na\"ive extension of two-party techniques, which would 
require
total communication of
$O(|V|^2 |A|^2)$ where $V$ is the set of vertices and $A$ the parties.
We achieve total communication of $O((|A| + |V|) |A| |V|)$. All participants are equal---there is no centralisation; and all parties are protected by edge DP. 
While we reuse the  \cite{roohi2019differentiallyprivate} subset release two-stage sampler, we offer new analysis. We prove in Proposition~\ref{prop:disjoint-exp-utility} that it can be distributed without privacy/accuracy loss, and establish a new high-probability utility bound on EBC based on this mechanism's release (Theorem~\ref{thm:utility-1}). 

\section{Related Work}
In most prior work on
differentially-private graph processing, the computation is performed by a trusted authority with complete knowledge of the whole network. Only the output, not the intermediate computations, must preserve privacy~\cite{hay2009accurate,bhaskar2010discovering,karwa2011private,zhang2015private,day2016publishing,mulle2015privacy,shen2013mining,kasiviswanathan2013analyzing,raskhodnikova2015efficient}. There is considerable work on distributed differential privacy~\cite{dwork2006our,chen2012towards,mohammed2013secure}, where queries are distributed or decomposed into sub-queries. 
However there is far less work in our distributed privacy model, in which even the intermediate communication should preserve differential privacy. This privacy model is mostly related to distributed graphs where parties seek joint computation of statistics of the graph. The most closely related work is~\cite{roohi2019differentiallyprivate}, which derives an edge-DP algorithm for EBC, but only for two networks. The algorithm allows two mutually-distrusting parties to jointly compute the EBC privately using the exponential and Laplace mechanisms to maintain differentially-private intermediate computations and communications. 
However their work assumes the first party acts as a centraliser that does not share the final result.
It is thus not directly applicable to our setting.

\section{Preliminaries}  \label{sec:preliminaries}

\subsection{Egocentric Betweenness Centrality}
Proposed by  \cite{everett2005ego} as a
way of measuring a node's importance by looking only at its \emph{ego network}, the set of nodes directly connected to it.  To compute the EBC of node $a$, we count, for each pair of $a$-neighbouring nodes $(i,j)$, what fraction of shortest paths between them pass through $a$.  Since we consider only paths within $a$'s ego network, only paths of length two are relevant.   We count zero for any pair $(i,j)$ that is directly connected, because these two nodes do not rely on $a$ at all.

\begin{definition}[\cite{everett2005ego}]\label{def2.2}
\term{Egocentric betweenness centrality} of node $a$ in simple undirected graph $(V,E)$ is defined as 
\begin{eqnarray*}
\EBC{a} &=& \sum_{i,j\in N_a : A_{ij}=0, j>i}
\frac{1}{A^2_{ij}}\enspace,
\end{eqnarray*}
where $N_a=\{v\in V\mid \{v,a\}\in E\}$ denotes the neighbourhood or \term{ego network} of $a$, $A$ denotes the $(|N_a|+1)\times (|N_a|+1)$ adjacency matrix induced by $N_a\cup\{a\}$ with $A_{ij}=1$ if $\{i,j\}\in E$ and $0$ otherwise;  $A^2_{ij}$ denotes the
$ij$-th entry of the matrix square, guaranteed positive for all $i,j\in N_a$ since all such nodes are connected through $a$.
\end{definition}

\subsection{Differential Privacy on Graphs} 

We wish to protect the privacy of \emph{connections} in networks, which
are the edges in the network graph. 
Differential privacy (DP)~\cite{dwork2006calibrating} limits the differences in an algorithm's output distribution
on neighbouring databases, thus quantifying the information leakage about
the presence or absence of a particular record.
We use \term{edge privacy} \cite{hay2009accurate}, because we
wish to control the attacker's ability to make inferences about the
presence of individual edges. 
As graphs on identical node sets, two databases are neighbours if they differ by exactly one edge.

Our databases are simply adjacency matrices in which an element $A_{ij}$ is 1 if there is an edge from $i$ to $j$ and zero otherwise.
Equivalently, these can be considered as sequences of bits: elements of $\{0,1\}^n$ (where $n$ is the number of nodes in the network choose two). 
Formally, two databases $D,D'\in\{0,1\}^n$  are termed \term{neighbouring} (denoted $D\sim D'$) if there exists exactly one $i\in [n]$ such that $D_i\neq D'_i$ and $D_j=D'_j$ for all $j\in [n]\backslash \{i\}$.
In other words, $\|D - D'\|_1=1$.

\begin{definition}
For $\epsilon>0$, a randomised algorithm on databases or \term{mechanism} $\mathcal{A}$ is said to preserve \term{$\epsilon$-differential privacy} if for any two neighbouring
databases $D, D'$, and for any measurable set $R\subseteq\mathrm{Range}(\mathcal{A})$,
\begin{eqnarray*}
\Pr\left(\A(D) \in R\right) &\leq& \exp(\epsilon)\cdot \Pr\left(\A(D') \in R\right)\enspace.
\end{eqnarray*}
\end{definition}

In this paper, we employ several generic mechanisms from the differential privacy literature including Laplace~\cite{dwork2006calibrating} for releasing numeric vectors, and the exponential~\cite{mcsherry2007mechanism} for private optimisation. We also use a recent subset release mechanism~\cite{roohi2019differentiallyprivate} which leverages the exponential mechanism, and develop its theoretical utility analysis.

\begin{lemma}[\cite{roohi2019differentiallyprivate}]\label{lem:subset}
Consider a publicly-known set $V$ and a privacy-sensitive subset $R^\star\subseteq V$. 
The exponential mechanism run with quality function $q(R^\star,R)=|R\cap R^\star|+|\overline{R\cup R^\star}|$ and $\Delta q=1$ preserves $\epsilon$-DP.  Algorithm~\ref{alg:forward} (see Appendix) implements this mechanism, running in $O(|V|)$ space and time.
\end{lemma}

\section{Problem Statement}\label{sec:problem}

We have $|A|$ participating parties   
$A=\{\alpha_1,\ldots,\alpha_{|A|}\}$, each representing
a \linebreak[4]telecommunications service provider.
They control a global communication graph $(V,E)$ whose nodes are 
partitioned into $|A|$ (disjoint) sets one per 
service provider s.t. $V_\alpha$ contains the nodes of party $\alpha$.
Every customer is represented as a node that belongs to one and only one service provider; pairs of customers who have had some communication (\eg a phone call, SMS or email) are edges. 

We will often equivalently represent edge sets as adjacency matrices (or flattened vectors) with elements in $\{0,1\}$.

We write $E_{\alpha,\alpha}$ for the set of edges in $(V,E)$ between nodes in $V_{\alpha}$---these are communications that happened entirely within $\alpha$. Similarly, $E_{\alpha,\beta}$ are the  edges with one node in $V_{\alpha}$ and the other in $V_\beta$---these represent communications between two service providers. 
Set $E = \bigcup_{\alpha, \beta\in A}
E_{\alpha,\beta}$ is the disjoint union of all such edge sets.

We assume that all nodes are known to all parties, but that each party learns only about the edges that are incident to a node in its network, including edges within its network. 

We wish to enable all parties to learn and publicly release the EBC of any chosen node $a$, while maintaining \emph{edge privacy} between all parties.  Without loss of generality we assume
$a \in V_{\alpha_1}$. We also denote by $V^-=V\backslash\{a\}$.
Before detailing a protocol for accomplishing this task, we must be precise about a privacy model.

\begin{problem}[Private Multi-Party EBC] \label{prob:pebc}
Consider a simple undirected graph\linebreak[4]($\bigcup_{\alpha\in A} V_\alpha$, $\bigcup_{\alpha, \beta\in A}E_{\alpha,\beta}$) partitioned by parties $A=\{\alpha_1,\ldots,\alpha_{|A|}\}$ as above, and an arbitrary node $a\in V_{\alpha_1}$. The problem of \term{private multi-party egocentric betweenness centrality} is for the parties $\alpha$ to collaboratively approximate \EBC{a} under  assumptions that:
\begin{enumerate}[{A}1.]
    \item All parties $\alpha\in A$ know the entire node set $\bigcup_{\alpha\in A} V_\alpha$; \label{ass:nodes}
    \item Each party $\alpha \in A$ knows every edge incident to nodes within its own network, \textit{i.e. }  $\bigcup_{\beta\in A} E_{\alpha,\beta}.$
    \label{ass:edges}
    \item The computed approximate \EBC{a} needs to be available to all of parties. \label{ass:ebc}
\end{enumerate}

The intermediate computation must protect $\eps$-differential edge privacy of each party from the others.  
We seek solutions under a fully adversarial privacy model: irrespective of whether other parties 
follow the protocol, the releases by party $\alpha$ protect its edge differential privacy. 
(Of course a cheating participant can always release information about edges it already knows, which may join another
network.)

Furthermore, the \emph{output} must protect $\eps$-differential privacy of the edges.
In \cite{roohi2019differentiallyprivate} the final EBC could be revealed only to the party who made the query.  In this paper, the final EBC is $\epsilon$-differentially private and can be released safely to anyone.

\end{problem}

\section{Multi-Party Private EBC}\label{sec:multiPartyProtocol}
We describe three algorithms: \SubsetRelease, \PrivatePathCount, and \linebreak[4]\PrivateReciprocateandSum, which are privacy-preserving versions of Steps~\ref{step:egonetwork}--\ref{step:reciprocatesum1} of Protocol~\ref{prot:NonPrivate} (See Appendix). These then combine  to produce \PrivateEBC, a differentially-private version of the whole protocol.

\subsection{Private Ego Network Broadcast }
Each party $\alpha$ runs \SubsetRelease, Algorithm~\ref{alg:forward} (see Appendix) with its share $R^\star_\alpha$  of $a$'s ego network. It broadcasts the output $R_\alpha$---the \linebreak[4]approximation of $R^\star_\alpha$.

\SubsetRelease uses the exponential mechanism to privately optimise a particular quality function (Lemma~\ref{lem:subset}) that encourages a large intersection between $R^\star_\alpha$ and release $R_\alpha$, along with a minimal symmetric set difference. As each party runs this mechanism relative to its own node set, it operates its own quality function defined relative to $R^\star_\alpha$ (see Proposition~\ref{prop:disjoint-exp-utility} for the formal definition). 
We observe a convenient property of the quality functions run by each party: they sum up to the overall quality function if the ego party was to run \SubsetRelease in totality. This permits proof (see the proof of proposition~\ref{prop:disjoint-exp-utility} in~\ref{subsec:proofOfExpUtility}) that this simple distributed protocol for private ego network approximation exactly implements a centralised approximation. \textbf{\textit{There is no loss to privacy or accuracy due to decentralisation.}}

\begin{proposition}
\label{prop:disjoint-exp-utility}
Consider parties $\alpha\in A$ running \SubsetRelease with identical budgets $\epsilon_1>0$
and quality functions $q_\alpha(R)=|R\cap R^\star\cap V_\alpha|+|\overline{R\cup R^\star}\cap V_\alpha|$, on their disjoint shares $R^\star_\alpha=R^\star\cap V_\alpha$ to produce disjoint private responses $R_\alpha\subseteq V_\alpha$. Then 
$R=\cup_{\alpha\in A}R_\alpha$ is distributed 
as \SubsetRelease run with $\epsilon$, quality function $q(\cdot)$, on the combined $R^\star$ in $V$. Consequently the individual $R_\alpha$ and the combined $R$, each preserve $\epsilon_1$-DP simultaneously.
\end{proposition}

\subsection{Private Path Count} \label{sec:backward}

\begin{algorithm}[tb]
 \caption{\PrivatePathCount}\label{alg:privatepathcount}
 \begin{algorithmic}[1]
 \renewcommand{\algorithmicrequire}{\textbf{Input:}}
 \REQUIRE 
   ego node $a\in V_{\alpha_1}$ (remember, by assumption, $\alpha_1$ contains $a$); execution party $\alpha \in A$; true node set $R^\star_\alpha$; for each $\beta \in A$, edge set $E_{\alpha,\beta}$ and private node set $R_\beta$; $\epsilon_{2}, \Delta_{2}>0$
  \ENSURE{A vector of noisy counts, indexed by endpoints $\{i,j\}$ with $i<j$, of the total number of nodes $k$ in $R^\star_\alpha$ that are connected to both $i$ and $j$.}
 \IF{$\alpha=\alpha_1$}
   \STATE $R^\star_\alpha \longleftarrow R^\star_\alpha \cup \{a\}$
 \ENDIF \\
 \STATE $R_A\longleftarrow \bigcup_{\beta\in A} R_{\beta}$
 \FOR{$i\in R_A$}
   \FOR{$j \in R_A$ with $i<j$}
     \STATE $K \longleftarrow \left\{k\in R^\star_\alpha \mid \{i,k\}, \{k,j\} \in \bigcup_{\beta\in A} E_{\alpha,\beta} \right\} $
     \STATE $T^{ij}_\alpha \longleftarrow |K| + \mathrm{Lap}(2 \Delta_{2}/\epsilon_{2})$
   \ENDFOR
 \ENDFOR 
 \RETURN{$\mathbf{T_\alpha}$}
\end{algorithmic}
\end{algorithm}

Each party $\alpha$ runs Algorithm~\ref{alg:privatepathcount}, using the $R_\beta$'s received from each other party in the previous step.
Party $\alpha$ counts all the 2-paths where the intermediate node is in $R_\alpha$.
For each node pair $(i,j)$ with $i<j$, $\alpha$ will send the 2-path count $T^{i,j}_\alpha$ to the party that 
contains node $i$, just like the non-private version of the protocol.
But first, in order to privatise this vector of counts, Laplace noise is added to the two-path counts
according to the sensitivity in the following lemma proved in Appendix~\ref{subsec:proofOfFSensitivity}, thereby preserving $\epsilon_2$-DP in this stage's release.
\begin{algorithm}[tb]
\caption{\PrivateReciprocateandSum}\label{alg:reciprocateAndSum}
 \begin{algorithmic}[1]
 \renewcommand{\algorithmicrequire}{\textbf{Input:}}
 \REQUIRE ego node $a\in V_{\alpha_1}$; execution party $\alpha \in A$; for each $\alpha\leq \beta \in A$, edge set $E_{\alpha,\beta}$ and private node set $R_\beta$; for each $\beta\in A$, noisy counts $\mathbf{T}_\beta$; $\epsilon_{3}, \Delta_{3}>0$  \\
 \STATE $R\longleftarrow \bigcup_{\beta>\alpha} R_\beta \cup R^\star_\alpha$
 \STATE $E_\alpha\longleftarrow\bigcup_{\beta\geq\alpha} E_{\alpha,\beta}$
 \STATE $S_\alpha \longleftarrow 0$
 \FOR{$i\in R^\star_\alpha$}
   \FOR{$j\in R$ with $i<j$}
     \IF{$\{i,j\}\notin E_\alpha$} \label{line:reciprocateIf}
       \STATE $T \longleftarrow \sum_{\gamma\in A} T_\gamma^{ij}$
       \STATE $S_\alpha \longleftarrow S_\alpha + \left(\lfloor\max\{0, T\}\rfloor + 1\right)^{-1}$
     \ENDIF
   \ENDFOR
 \ENDFOR
 \STATE $S_\alpha \longleftarrow S_\alpha +\mathrm{Lap}(2 \Delta_{3}/\epsilon_{3})$
 \RETURN $\mathbf{S_\alpha}$
\end{algorithmic}
\end{algorithm}

\begin{algorithm}[tb]
\caption{\PrivateEBC}\label{alg:privateEBC}
 \begin{algorithmic}[1]
 \renewcommand{\algorithmicrequire}{\textbf{Input:}}
\REQUIRE (Public) ego node $a\in V_{\alpha_1}$; ordered set of parties $A$; node sets $V_\alpha$ for $\alpha\in A$; parameter vectors $\eps,\Delta\succ 0$. \\
\REQUIRE (Private) for each $\alpha, \beta \in A$, edges $E_{\alpha,\beta}$, nodes $R^\star_\alpha$; \\ 
\FOR{$\alpha \in A$ in parallel}
    \STATE{Party $\alpha$ does:}
     \IF{$\alpha=\alpha_1$} 
     \STATE $V=V_\alpha \backslash\{a\}$
   \ELSE
     \STATE $V=V_\alpha$
   \ENDIF
    \STATE{$R_\alpha \leftarrow \SubsetRelease(V,R^\star_\alpha, \epsilon_1)$}
    \STATE{Broadcast $R_{\alpha}$}
    \STATE{$\mathbf{T_\alpha} \leftarrow \PrivatePathCount(\alpha,E_{\alpha \beta},R_\beta,\epsilon_{2},\Delta_{2})$}
    \FORALL{$i,j \in V$ with $i < j$}
        \STATE{Send $T^{i,j}_{\alpha}$ to the Party $\beta$ s.t. $i \in V_{\beta}$}
    \ENDFOR
    \STATE{$S_\alpha \leftarrow \ReciprocateandSum(a, \{E_{\alpha,\beta}, R_\beta | \beta > \alpha\},\epsilon_3)$.}  
    \COMMENT{Party $\alpha$ reciprocates and sums only paths with $\beta \geq \alpha$.}
    \STATE{Broadcast $S_\alpha$}
    \STATE{$\PEBC{a} \leftarrow \sum_{\alpha \in A} S_\alpha$}
    \STATE{Return $\PEBC{a}$}
\ENDFOR
 
\end{algorithmic}
\end{algorithm}

\begin{lemma}
\label{lem:fsensitivity}
Let query $f$ denote the vector-valued non-private response $\mathbf{T}_\alpha$ of party $\alpha$ in Algorithm~\ref{alg:privatepathcount}. 
The $L_1$-global sensitivity of $f$ is upper-bounded by $\Delta f=2|R_A|$.
\end{lemma}

\subsection{Private Reciprocate and Sum}

Every party $\alpha$ receives noisy counts from \PrivatePathCount and for any pairs $i<j$ where $i\in R^\star_\alpha$ and $j\in \bigcup_{\beta >\alpha} R_\beta \cup R^\star_\alpha$, that are believed by $\alpha$ to be disconnected, $\alpha$ increments the received $T^{i,j}$ by the number of incident 2-paths. 
Each party then reciprocates the summation of the counts. In this algorithm, each party may replace noisy $R_{\alpha}$ with true $R^\star_{\alpha}$. This optimises utility at no cost to privacy:{ counts $T^{ij}$ for $i,j\in R_\alpha\backslash R^\star_\alpha$ are discarded.}  This is safe to do, since the Laplace mechanism already accounts for changes in $R^\star_\alpha$. 
The Laplace noise is utilised to privatise the reciprocated sum $S_\alpha$ to $\epsilon_3$-DP, calibrated by  sensitivity as bounded next, with proof can be found in Appendix~\ref{subsec:proofOfFPrime}. 
\begin{lemma}  \label{lem:f-primeSensitivity}
Let query $f'$ denote the reciprocate and sum over 2-paths with intermediate point in $\bigcup_\beta R_\beta$ while the nodes $i<j$ are not connected and $i\in R^\star_\alpha$ and $j\in \bigcup_{\beta<\alpha} R_\beta \cup R^\star_\alpha$.
Then the $L_1$-global sensitivity of $f'$ is upper-bounded by $\Delta f'= (\lfloor\max\{0, T\}\rfloor + 1)^{-1}\leq 1$ irrespective of party.
\end{lemma}

\paragraph*{Communication complexity}
Ego Network Broadcast requires each party to send to each other party $|V|$ bits of length 1 that shows the node is present or not, hence a total of $O(|A|^2 |V|)$.  Private Path Count sends, for each node $i$, up to $|V|$ messages $T^{i,j}$ from each party to the owner of node $i$.  The pathcounts $T^{ij}$ are at most $|V|$, so the total size is $O(|A| |V|^2)$.
Finally, Reciprocate and Sum requires every participant to send each other one message: $O(|A|^2)$.  Hence the total communication complexity is $O((|A| + |V|) |A| |V| )$.

\subsection{\PrivateEBC: Putting it All Together}

After the parties have run the protocol phases, namely \SubsetRelease, \PrivatePathCount and \PrivateReciprocateandSum, they must finally complete the computation of the private EBC.  Algorithm~\ref{alg:privateEBC} depicts \PrivateEBC  orchestrating the high-level protocol thus far, and then  adding the received $S_\alpha$ to compute final EBC. 

\begin{theorem}
\PrivateEBC preserves $(\epsilon_1+\epsilon_2+\epsilon_3)$-DP for each party.
\end{theorem}

\begin{remark}
While we have used uniform privacy budgets across parties, our analysis immediately extends to custom party budgets. 
\end{remark}
\captionsetup[subfigure]{subrefformat=simple,labelformat=simple,listofformat=subsimple}
\renewcommand\thesubfigure{(\alph{subfigure})}
\begin{figure*}[ht]
\centering
\subfloat[Median relative error of the 60 random nodes with $\epsilon=$0.1 to  7, Facebook, Enron and PGP data set, for three parties.]{\label{fig:a1}\includegraphics[width=0.490\textwidth, keepaspectratio]{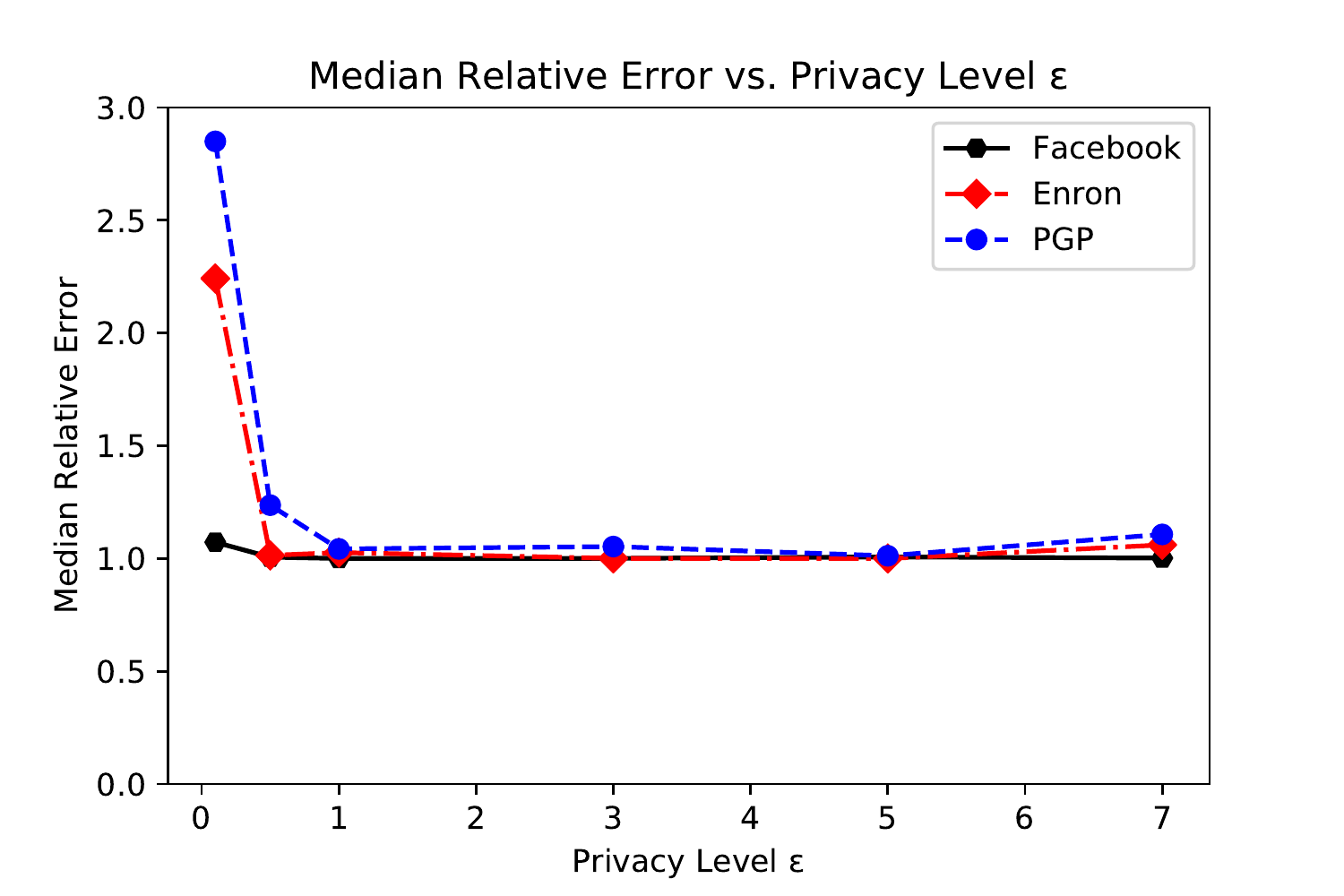}
}
\hfill
\subfloat[Median Relative error of 120 nodes with $\epsilon=1$, for different number of parties for PGP.]{\label{fig:b1}\includegraphics[width=0.490\textwidth, keepaspectratio]{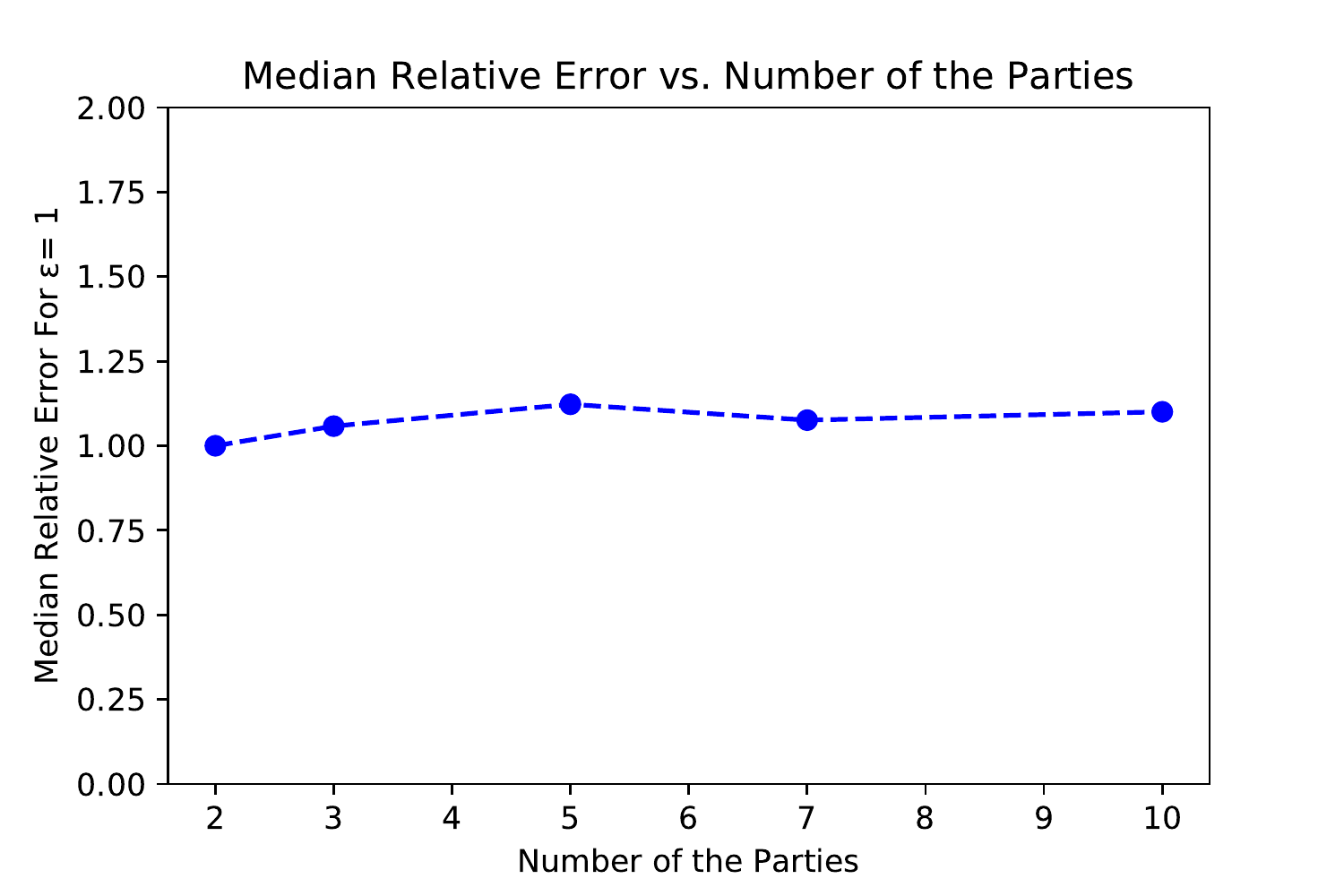}
}
\caption{Utility of Private EBC for Facebook, Enron and PGP data sets.}
\label{fig_sim1}
\end{figure*}
\section{Utility Bound} \label{sec:utilityBound}

In this section we develop a utility analysis of privacy-preserving betweenness centrality, noting that no previous theoretical analysis has been performed including in the two-party case~\cite{roohi2019differentiallyprivate}. 
Our analysis focuses on a utility bound on EBC resulting from the subset release mechanism run to privatise the ego network. 
We abuse notation with $q(R)=q(R^\star,R)$ referring to the quality function of the \SubsetRelease mechanism of Lemma~\ref{lem:subset} with dependence on the private $R^\star$ made implicit; likewise for the quality functions run by each party in the decentralised setting. The technical challenge is in leveraging the following well-known utility bound on the exponential mechanism, which only establishes high-probability near-optimal quality.

\begin{corollary}\label{cor:exp-mech-combined}
Consider parties $\alpha\in A$ each running \SubsetRelease concurrently with budgets $\epsilon>0$ and quality functions $q_\alpha(R)=|R\cap R^\star\cap V_\alpha|+|\overline{R\cup R^\star}\cap V_\alpha|$ on their disjoint shares $R_\alpha^\star=R^\star\cap V_\alpha$ to produce disjoint responses $R_\alpha\subseteq V_\alpha$. Then the consequent high-probability quality bound of Lemma~\ref{lem:mcSherry7} (see Appendix) holds for random combined response $R=\cup_\alpha R_\alpha$.
\end{corollary}

Our first step is to relate EBC error on a released $R$ to the quality $q(R)$. We organise differences in EBC by reciprocal 2-path count terms, enumerating shared and unshared such terms between private and non-private EBCs. As these terms and their differences are bounded by one, the task reduces to measuring differences in ego network cardinalities, Lemma~\ref{Lemm:utility_1} (see Appendix). Conveniently, this is also the goal of our quality score function. 
We now use Lemma~\ref{Lemm:utility_1} with our lifted exponential utility bound Proposition~\ref{prop:disjoint-exp-utility} to bound EBC error.
The previous lemma is agnostic to the number of parties---the released $R$ might be produced in its entirety by the ego node's party through a single call to \SubsetRelease, or it could be the disjoint union of multiple calls to \SubsetRelease by each party. Likewise our lifted bound on high-probability quality also holds as if $R$ is produced centrally. As such the proof---found in the Appendix---of the following result may proceed as if this is the case.

We prove the following high-probability utility bound in the Appendix~\ref{subsec:proofOfUtilityThm}.
\begin{theorem}\label{thm:utility-1}
Consider privacy budget $\epsilon>0$, true ego network $R^\star$ and $t>|R^\star|^2/2$.
And suppose that each party $\alpha\in A$ runs \SubsetRelease with 
budget $\epsilon$, 
quality function $q_\alpha(R)=|R\cap R^\star\cap V_\alpha|+|\overline{R\cup R^\star}\cap V_\alpha|$, on their disjoint share $R^\star_\alpha=R^\star\cap V_\alpha$ to produce disjoint private response $R_\alpha\subseteq V_\alpha$. Then $EBC_1$ produced from $R=\cup_{\alpha\in A}R_\alpha$ incurs
error relative to non-private EBC run on non-private $R^\star$, 
upper bounded as
$|EBC - EBC_1| \leq t, $
with probability at least:  $ 1 - \exp\left(-\epsilon (\sqrt{2 t} - |R^\star|) / 2\right) 2^{|V^-|}\ .$
\end{theorem}

\begin{remark}
\label{remark:non-vacuous}
The bound of Theorem~\ref{thm:utility-1} can make meaningful predictions (\ie is non-vacuous).
For example \textbf{a modest  privacy budget of 2.1 is sufficient to guarantee reasonable relative error 3 w.h.p 0.999} for a large ego network spanning half an (otherwise sparse) graph. Similar  relative error (for end-to-end private EBC) at similar privacy budgets occurs in experiments on real, non-sparse networks below.
Further analysis can be found in Appendix~\ref{subsec:nonVac}.
\end{remark}

\begin{figure*}[ht]
\centering
\subfloat[Time of computing 60 random nodes with $\epsilon=$0.1 to 7, Facebook data , for three parties. ]{\label{fig:b}\includegraphics[width=0.32\textwidth, keepaspectratio]{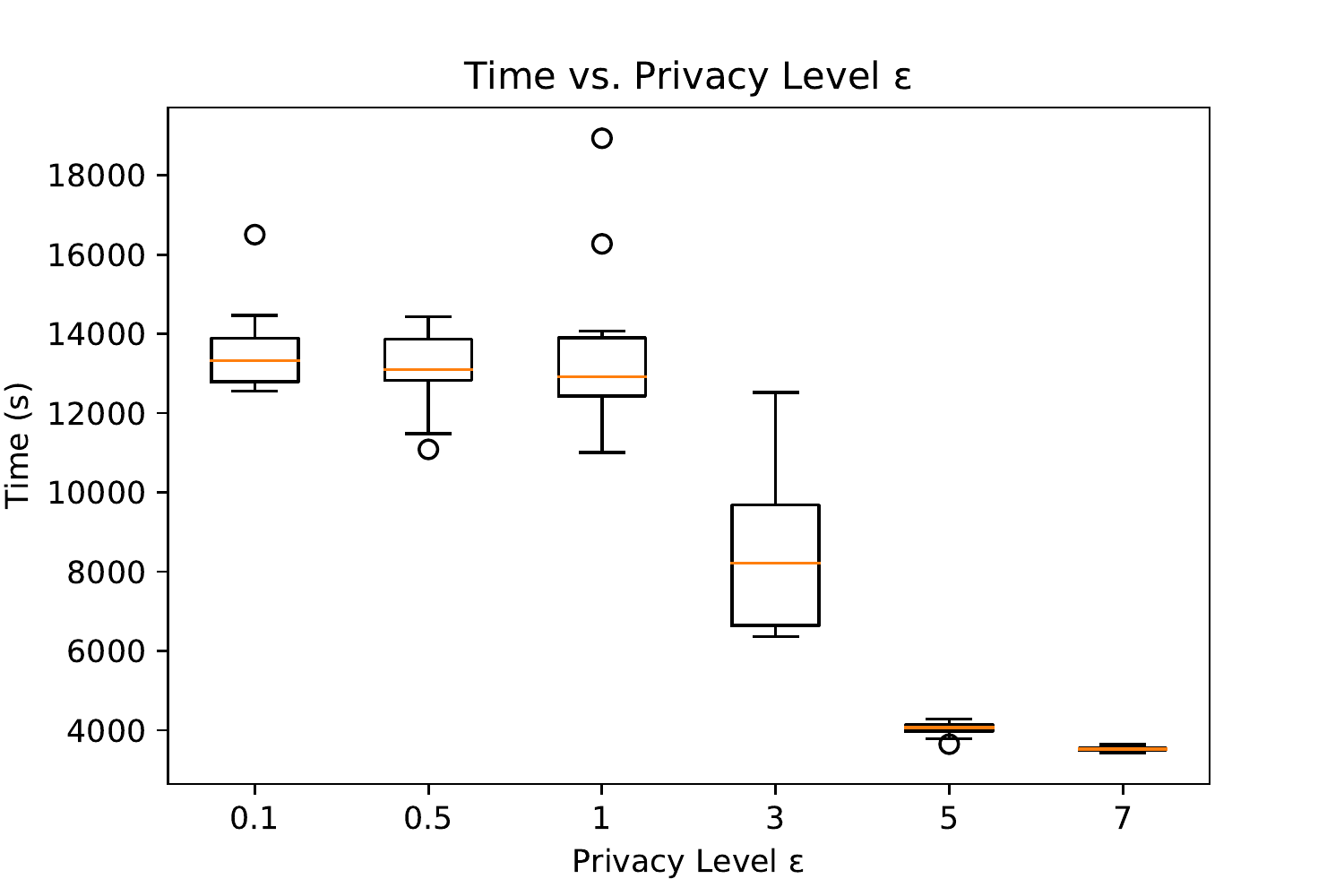}
}
\hfill
\subfloat[Time of computing 60 random nodes with $\epsilon=$0.1 to 7, Enron data set , for three parties. ]{\label{fig:e}\includegraphics[width=0.32\textwidth, keepaspectratio]{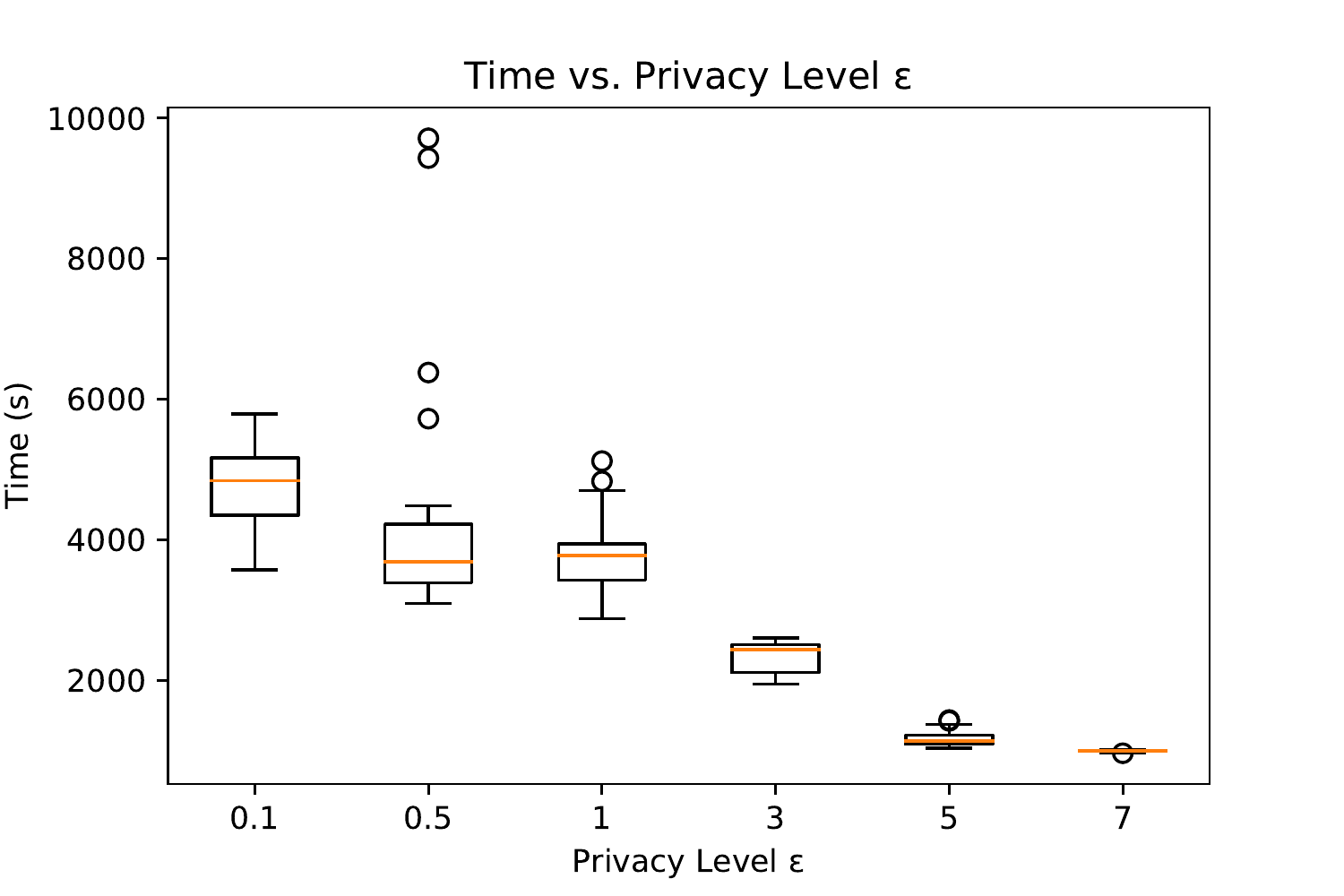}
}
\hfill
\subfloat[Time of computing 60 random nodes with $\epsilon=$0.1 to 7, PGP data set, for three parties. ]{\label{fig:h}\includegraphics[width=0.32\textwidth, keepaspectratio]{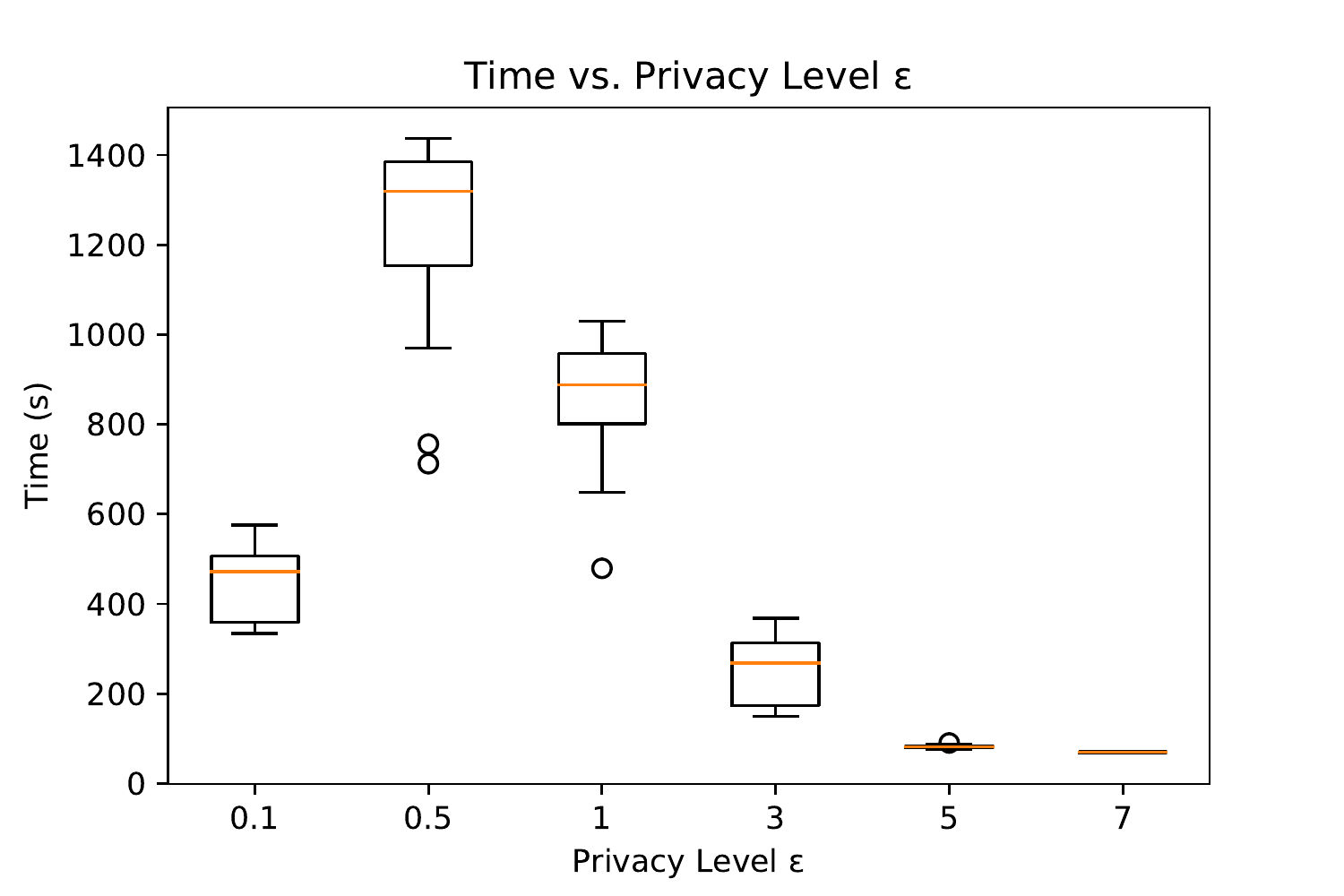}
}
\hfill
\subfloat[Relative error of 60 nodes with different degrees for $\epsilon=$1, Facebook data set.]{\label{fig:c}\includegraphics[width=0.32\textwidth, keepaspectratio]{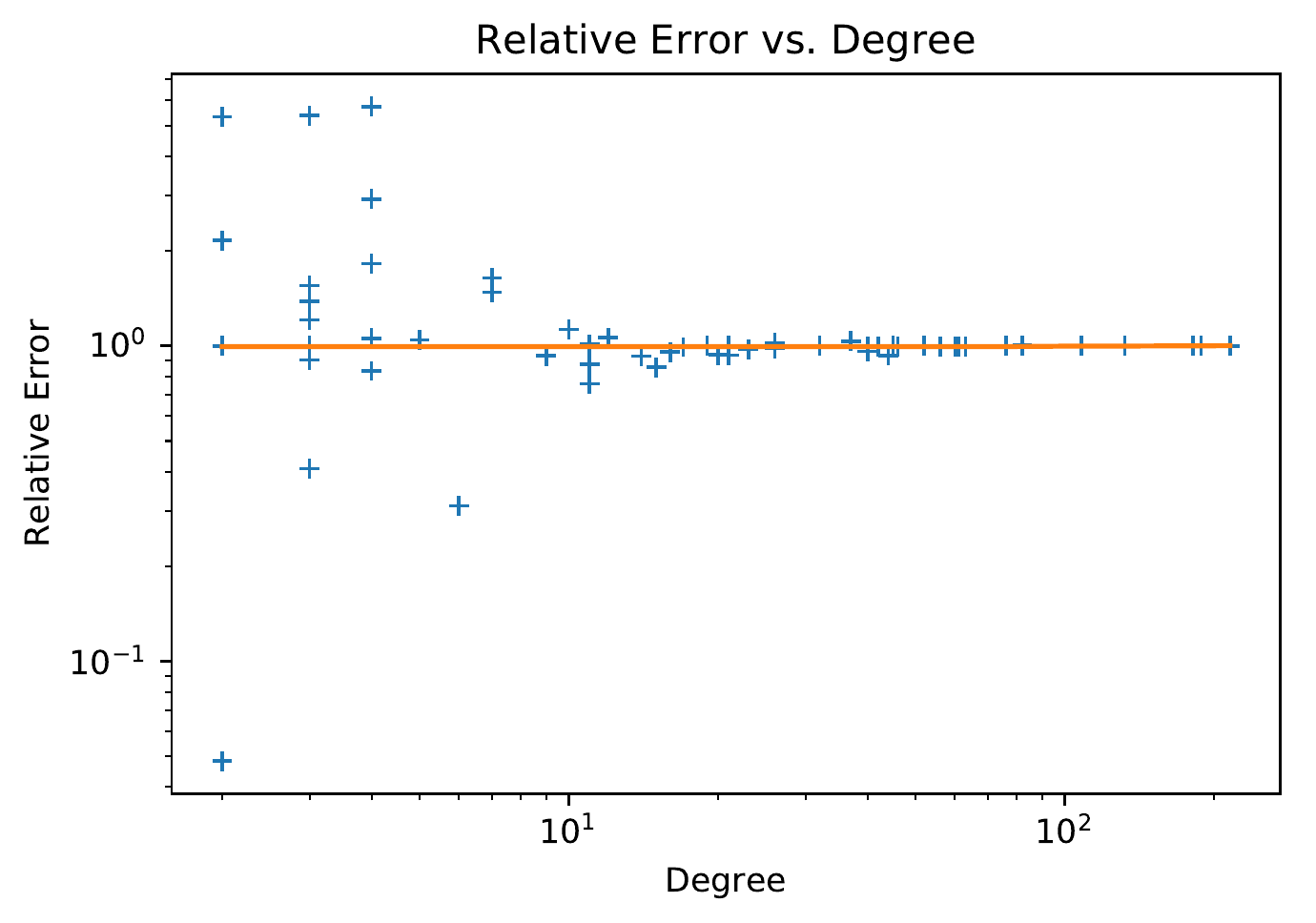}
}
\hfill
\subfloat[Relative error of 60 nodes with different degrees for $\epsilon=$1, degree, Enron data set.]{\label{fig:f}\includegraphics[width=0.32\textwidth, keepaspectratio]{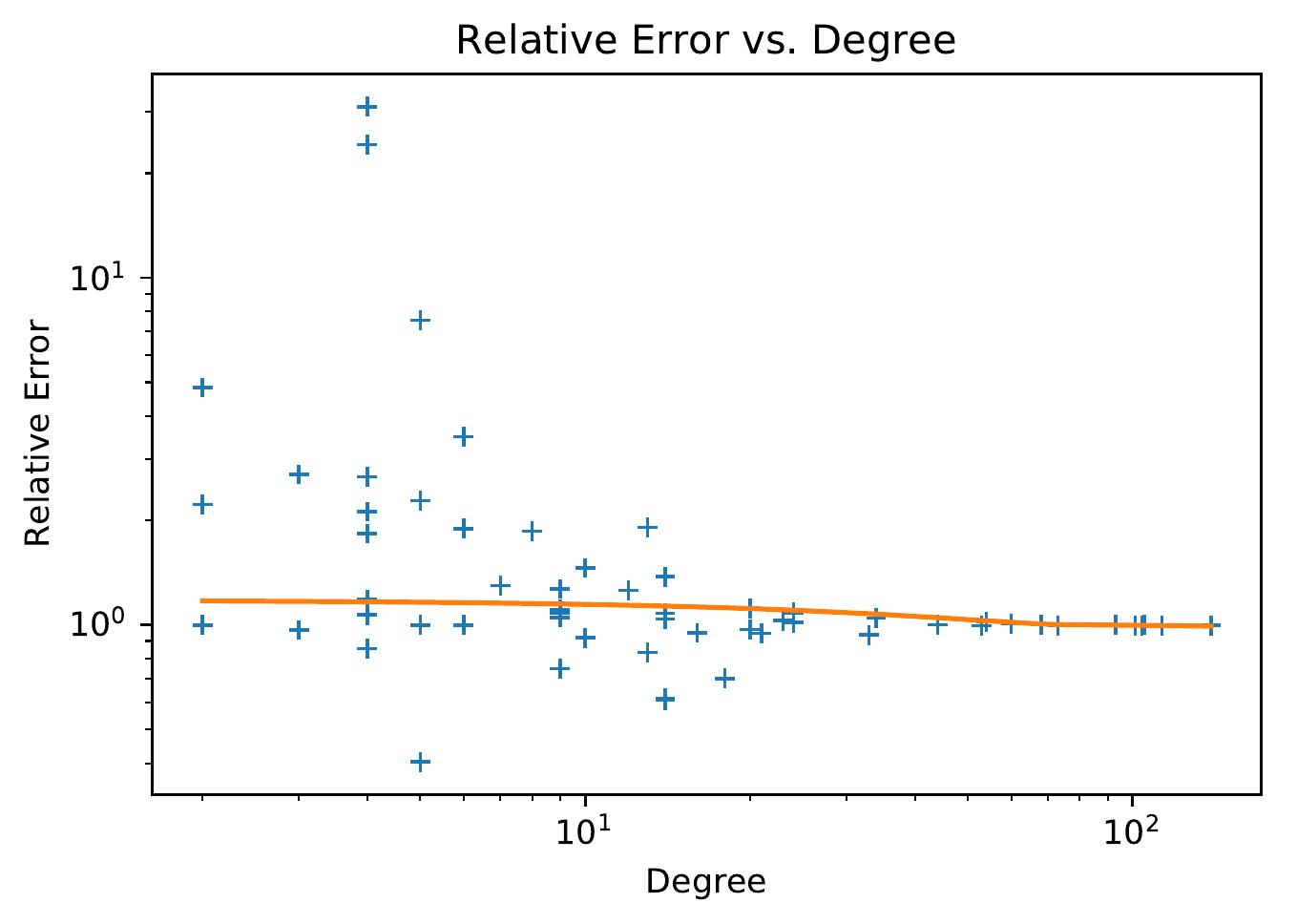}
}
\hfill
\subfloat[Relative error of 60 nodes with different degrees for $\epsilon=$1, PGP data set.]{\label{fig:i}\includegraphics[width=0.32\textwidth, keepaspectratio]{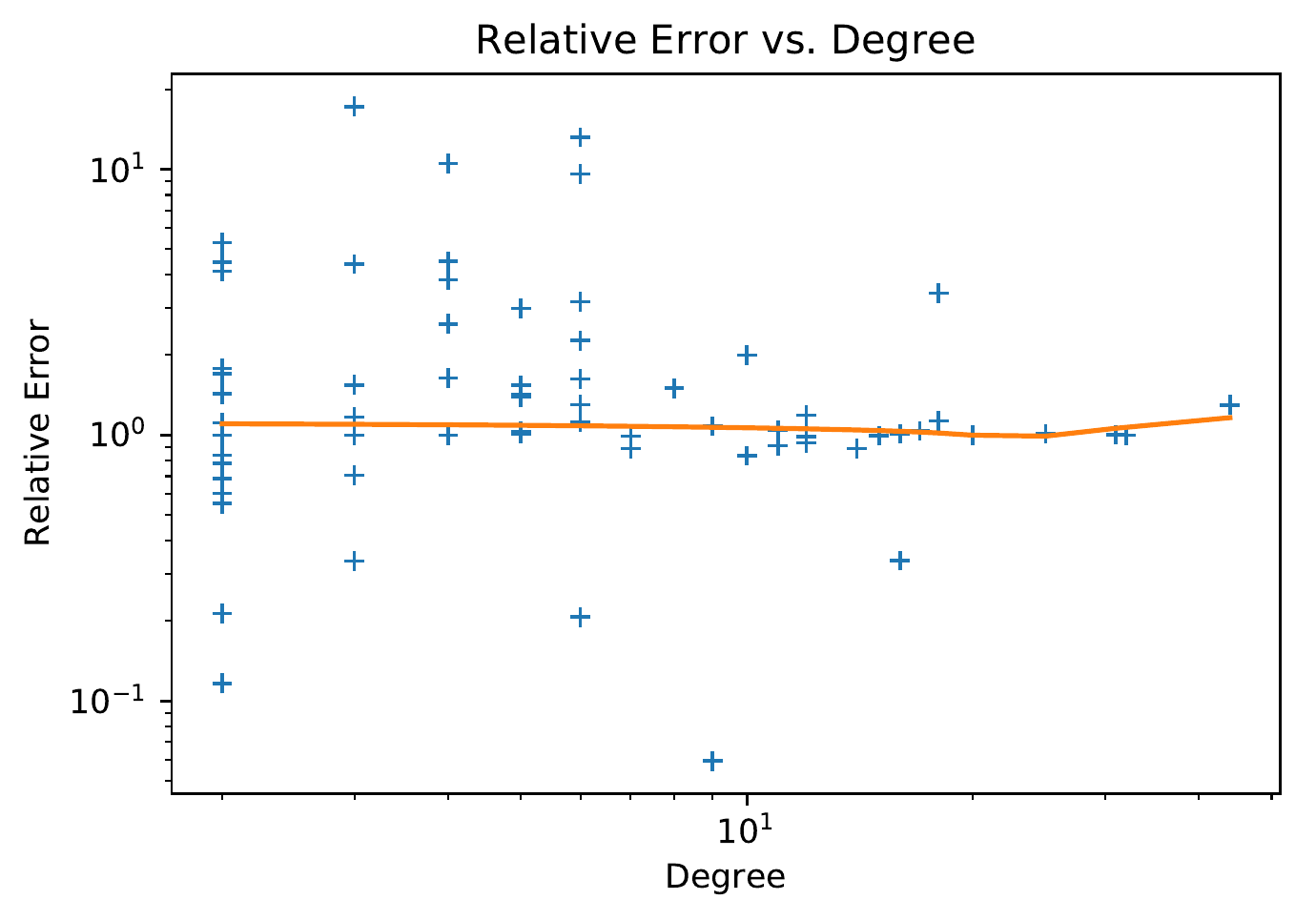}
}
\caption{Timing results and effect of degree for Facebook, Enron and PGP data sets.}
\label{fig_sim}
\end{figure*}
\section{Experimental Setup}
In order to validate the utility and privacy of \PrivateEBC, we experimented with three different graphs on Facebook friendships with 63,731 nodes and 817,035 edges\footnote{Institute of Web Science and Technologies at the University of Koblenz–Landau: The Koblenz network collection (2018), \label{footnote1}}, the Enron email network with 36,692 nodes and 183,831 \linebreak[4]edges\footnote{Stanford University: Stanford large network dataset collection (2009)}and Pretty Good Privacy (PGP) with 10,680 nodes and 24,316 edges\textsuperscript{\ref{footnote1}}. We employ uniform random sampling in order to partition the graphs into multiple disjoint parties while keeping the structure of the graph intact. 
In addition to evaluations on three parties across datasets, we also validated utility across 2, 3, 5, 7 and 10 parties on the PGP data set. 
The experiments were run on a server with $2\times28$ core Xeon's (112 threads with hyper threading) and 1.5 TB RAM, using Python~3.7 without parallel computations. We employed the \texttt{Mpmath} arbitrary precision library for implementing inverse transform sampling (Algorithm~\ref{alg:itsampler} in the Appendix) and set the precision to 300 bits.
We use relative error between true EBC and private EBC---the lower the relative error the higher the utility. Any errors around 1 or 2 are considered practical as they signify EBCs within the same order of magnitude. We ran the experiment 60 times for each chosen value $\epsilon$ by choosing the target ego nodes randomly and robustly aggregating the relative error by median. Throughout we set $\epsilon_1=\epsilon_2=\epsilon_3=\epsilon/3$.
\section{Results} \label{sec:results}
First, we demonstrate how \PrivateEBC utility varies with increasing privacy budget from 0.1 to 7, for three parties across each of three different graph datasets. The median relative error between real and private EBC represents utility. 
Figure~\subref*{fig:a1} displays the results for Facebook, Enron and PGP data sets, where median relative error decreases significantly when $\epsilon$ is increased to a strong guarantee of 1, and remains small for larger $\epsilon$. For strong privacy guarantee of $\epsilon = 0.5$, median relative error is usually $\approx 1$ for all three data sets. These results demonstrate that \textbf{\textit{\PrivateEBC achieves practical utility across a range of graph sizes and privacy levels.}}
Next we report utility at privacy $\epsilon=1$ for the number parties ranging over 2, 3, 5, 7 and 10. Every point in Figure~\subref*{fig:b1} shows the median relative error between private and real EBC across 120 randomly chosen nodes in the PGP data set.  
\textbf{\textit{Our results find insignificant degradation occurs to accuracy or privacy when growing the number of parties.}} 
\begin{remark}
While more parties means more calls to \ReciprocateandSum and \PrivatePathCount such that the scale of the second and third mechanisms' Laplace noise increases moderately, the major source of error, \SubsetRelease,

is not affected by the number of parties as proved in Proposition~\ref{prop:disjoint-exp-utility}.
\end{remark}
We report on timing analysis for \PrivateEBC as a function of privacy. Median computation time of 60 random ego nodes for $\epsilon$ budget from 0.1 to 7 is reported in Figures~\subref*{fig:b}, \subref*{fig:e} and \subref*{fig:c}, on Facebook, Enron and PGP data sets. Here total time overall decreases as privacy decreases (increasing $\epsilon$), while a small increase to runtime can be seen at very high levels of privacy (low but increasing $\epsilon)$ for Enron it is likely due to different behaviours in the protocol with increasing $\epsilon$. When the set difference of $R$ and $R^\star$ is small, the two-stage sampler generates small numbers of nodes in faster time.
However faster runtime with lower privacy dominates behaviour overall. 

Figures~\subref*{fig:c}, \subref*{fig:f}, \subref*{fig:i} show how the median relative error is changing by ego node degree. We report results on privacy budget $\epsilon=1$, which do not show significant dependence: In Facebook the median relative error is almost constant for different node degrees and in Enron and PGP for node degrees up to $10^2$, deviations are approximately 1\% and 0.5\% of the maximum relative error respectively. 

\section{Conclusion and Future Work}
This paper develops a new protocol for multi-party computation of egocentric betweeness centrality (EBC) under per-party edge differential privacy. We significantly improve on past work by extending to multiple parties, achieving very low communication complexity, theoretical utility analysis, the facility to release the private EBC to all parties. Experimental results demonstrate the practical accuracy and runtime of our protocol at strong levels of privacy.

For future work we hope to allocate differential privacy budgets per stage, by optimising utility bounds. 
We also intend to develop a network model that reflects a person's use of multiple media, so that the node set need not be disjointly partitioned, while the privacy of edges remains paramount.

\appendix
\section*{Appendix}
\section{Non-Private Multi-Party Protocol}\label{sec:nonprivate}
We first show how different parties can compute EBC without preserving privacy, but with special attention to efficiency (so as to improve on a na\"ive application of the two-party protocol of \cite{roohi2019differentiallyprivate}).
A party $\beta$ that contains a node $j$ can always count the number of 2-step paths through $j$, but it doesn't know which of the nodes adjacent to $j$ are in the ego-network of $a$ (except in some special cases).
So in order for $\beta$ to count the number of 2-paths in $a$'s ego network that pass through $j$, we require each other party $\alpha$ to tell $\beta$ which of its nodes are neighbours of $a$.  This is denoted by
$R^\star_{\alpha}$.

\begin{figure}[t!]
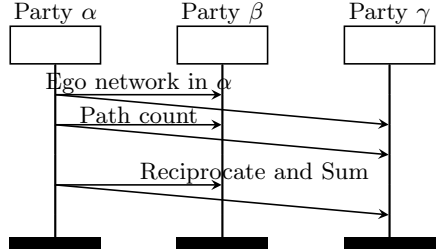

\centering
\vspace{-2em}
\drawframe{no}
\begin{msc}[small values, instance
distance=1cm,left environment distance=1.7cm,
right environment distance=1.7cm]{}

\declinst{usr}{Party $\alpha$}{}
\declinst{m1}{Party $\beta$}{}
\declinst{m2}{Party $\gamma$}{}

\mess
[label distance=0.07ex]{Ego network in $\alpha$}{usr}{m1}

\mess{}{usr}{m2}[1]

\nextlevel

\mess[label distance=0.07ex]{Path count}{usr}{m1}
\mess{}{usr}{m2}[1]

\nextlevel
\nextlevel
\mess[label distance=0.07ex][label position=above right]{Reciprocate and Sum}{usr}{m1}
\mess{}{usr}{m2}[1]
\end{msc}
\caption{EBC multi-party protocol for party $\alpha$, comprising three  messages per party. Visualised for three parties.} 
\label{fig:EBC}
\drawframe{yes}
\end{figure}

Recall that $N_a$
denotes the ego network of $a$ anywhere in the graph (\emph{not} including $a$ since the graph has no self-loops).  Figure~\ref{fig:EBC} summarises the following protocol.

\begin{protocol} All parties execute in parallel, waiting until they have received all messages from one step before commencing the next step.  Party $\alpha$ proceeds as follows:

\label{prot:NonPrivate}
\begin{enumerate}[i.]
   
\item {[EgoNetwork]} $\alpha$ broadcasts to every party the set $R^\star_\alpha$ of neighbours of $a$ contained within $\alpha$; \label{step:egonetwork}

\item {[PathCount]} For 
all nodes $i,j\in \bigcup_{\beta \in A} R^\star_\beta$ s.t. $i<j$, party
$\alpha$ computes $T^{i,j}_\alpha$, the number of 2-paths from $i$ to $j$ where the intermediate point $k\in R^\star_\alpha$ (irrespective of whether $i,j$ are directly connected). 
It sends $T^{i,j}_\alpha$ to the party $\hat{\beta}$ for which $i\in V_{\hat{\beta}}$;
\label{step:pathcount}
\item {[ReciprocateAndSum]} For every $i\in R^\star_\alpha$ and all $j>i$, $\alpha$ computes the total number of 2-paths between $i,j$ provided these nodes are disconnected: it sums $T^{i,j}_\beta$ for all $\beta\in A$. It then sets $S_\alpha$ to be the reciprocal of this sum and broadcasts this value to all parties; 
\label{step:reciprocatesum1}

\item $\alpha$ completes the computation of \EBC{a} as $\sum_{\alpha \in A} S_\alpha$.
\label{step:reciprocatesum2}
\end{enumerate}
Participants can easily tell when to move on to the next step.  At the end of Step~\ref{step:pathcount}, party $\alpha$ should have received $T^{i,j}_{\tilde{\beta}}$, from each other party $\tilde{\beta}$, for each node $i \in R^*_{\alpha}$, and each $j\in \cup_{\beta \in A} R^\star_\beta$ with $j>i$.  By the end of Step~\ref{step:reciprocatesum1} it should have received a broadcast value from all other parties.
\end{protocol}

\section{Privacy Disclosure}\label{rem:disclosure}
Now consider how the privacy of edges can be compromised in the first three steps of Protocol~\ref{prot:NonPrivate}.
\begin{enumerate}[i.]

\item When $R^\star_\alpha$ is broadcast in Step~\ref{step:egonetwork}, other parties  learn directly of all edges incident to $a$ in $\alpha$.

\item When $T^{ij}_{\alpha}$ is sent to $\hat{\beta}$ in Step~\ref{step:pathcount}, it reveals information about edges outside $\hat{\beta}$.
A worst case occurs for node $i$ when there is only one node $k \in \alpha$ connected to it.  Then $T^{ij}_{\alpha}$ reveals the existence of edge $j,k$ for all $j>i$.

\item When $\alpha$ broadcasts $S_\alpha$,
it reveals the connection status of edges within $\alpha$. In the worst case when there is just two nodes $i$ and $j$ in $\alpha$, an edge between them can change  $S_\alpha$ from a non zero value to zero. 
\end{enumerate}

\section{Proof of Proposition~\ref{prop:disjoint-exp-utility}}\label{subsec:proofOfExpUtility}
Observe that since the $V_\alpha$ form a disjoint partition of $V$, that for any $R_\alpha\subseteq V_\alpha$ we have that 
\begin{eqnarray*}
\sum_{\alpha\in A} q_\alpha(R_\alpha) &=& q\left(\bigcup_{\alpha\in A} R_\alpha\right)\enspace.
\end{eqnarray*}
Combining this with the independence of the concurrent executions of \SubsetRelease, we have 
that the joint density over their releases corresponds to 
\begin{eqnarray*}
p(r)
&=& \prod_{\alpha\in A} p(r_\alpha) \\
&\propto& \prod_{\alpha\in A} \exp\left( q_\alpha(r_\alpha)\cdot\epsilon_1\right) \\
&=& \exp\left(\epsilon_1 \sum_{\alpha\in A} q_\alpha(r_\alpha)\right) \\
&=& \exp\left( q(r)\cdot\epsilon_1\right)\enspace.
\end{eqnarray*}
Due to uniqueness of probability density normalisation, this proves the result.

\section{Proof of Lemma~\ref{lem:fsensitivity}}\label{subsec:proofOfFSensitivity}
Algorithm \PrivatePathCount is executed by all parties; the output of these computations is broadcast to other parties. As we need to preserve the privacy of each party's edges individually, we consider one (arbitrary) party $\alpha\in A$. The output of $\alpha$ is a vector of the counts of all $2$-paths connecting $i,j\in R_A=\bigcup_{\beta\in A} R_\beta$ with intermediate node $k\in R^\star_\alpha$. 
Adding or removing an edge from $E_{\alpha}=\bigcup_{\beta\in A} E_{\alpha,\beta}$, can worst-case change $2|R_A|$  elements of $\alpha$'s counts by one each. To see this, consider a very highly-connected node $k$ and the deletion of $\{i,k\}$ for any $i$: this reduces the counts for paths joining $i,j$ for all $j\in R_A$. Likewise if $i$ were also within party $\alpha$ and were highly connected then the edge removal would also reduce counts for paths joining $k,j$ for all $j\in R_A$. This proves that the sensitivity for part $\alpha$ running \PrivatePathCount is $2|R_A|$ irrespective of party $\alpha$.

\section{Proof of Lemma~\ref{lem:f-primeSensitivity}}\label{subsec:proofOfFPrime}

Once again we apply post-processing to previously sanitized outputs $R_\beta, T_\beta$. Consider any party $\alpha\in A$ and the effect of removing/adding an edge incident to a node of $\alpha$. At worst this will result in the condition $\{i,j\}\notin E_\alpha$ of line~\ref{line:reciprocateIf} of Algorithm~\ref{alg:reciprocateAndSum}, evaluating differently---for at most one
pair $\{i,j\}$. That is, the non-private sum of reciprocals $S_\alpha$ can be affected by the addition/removal of a single term $\left(\lfloor\max\{0, T\}\rfloor + 1\right)^{-1}$. Such a term is upper bounded by the reciprocal of lower bound on $\lfloor\max\{0, T\}\rfloor + 1\geq 1$. That is, the sensitivity, irrespective of executing party $\alpha$, is 1.

\section{Lemma~\ref{lem:mcSherry7}}
\begin{lemma}[Lemma 7 of \cite{mcsherry2007mechanism}]
\label{lem:mcSherry7}

Consider a centralised party running \SubsetRelease with budget $\epsilon>0$, quality function $q(\cdot)$ on $R^\star$. For $t>0$ let $S_t=\left\{R:q(R)>q(R^\star)-t\right\}$, and let $\mu$ be the uniform probability mass function on $\mathcal{P}(V^-)$.
Then 
$\Pr(\overline{S}_{2t}) \leq \exp(-\epsilon t)\mu(\overline{S}_{2t})/\mu(S_t)$.

\end{lemma}
\section{Lemma~\ref{Lemm:utility_1}}
\begin{lemma}\label{Lemm:utility_1}For any $R\subseteq V^-$ 
we may bound the additive EBC error  from using $R$ instead of non-private $R^\star$ according to the quality function applied to the random release:
$$|EBC-EBC_1|\leq \frac{1}{2}(|V|-q(R)+|R^\star|)^2$$
\end{lemma}
  \subsection{Proof of Lemma~\ref{Lemm:utility_1}}\label{sec:EBCerrorBound}

Let $T^{ij}, T_1^{ij}$ denote the number of 2-paths connecting $i,j$ within node-sets $R^\star$ and $R$ respectively. Our goal is to upper bound the quantity:
\begin{eqnarray*}
&& \left|\sum_{i,j \in R^\star} \frac{1\left[\left\{i,j\right\} \in E\right]}{T^{ij}}-\sum_{i,j \in R} \frac{1\left[\left\{i,j\right\} \in E\right]}{T_1^{ij}}\right| \\
&\leq& \sum_{i,j \in R^\star\cap R} \left|\frac{1\left[\left\{i,j\right\} \in E\right]}{T^{ij}}- \frac{1\left[\left\{i,j\right\} \in E\right]}{T_1^{ij}}\right| \\
&&+\; \sum_{i\in R\backslash R^\star, j\in R} \left|\frac{1\left[\left\{i,j\right\} \in E\right]}{T_1^{ij}}\right| \\
&&+\; \sum_{i\in R^\star\backslash R, j\in R^\star} \left|\frac{1\left[\left\{i,j\right\} \in E\right]}{T^{ij}}\right|\enspace,
\end{eqnarray*}
following from the triangle inequality and collection of terms with shared end-point nodes $i,j$. By cases: only when both end points are elements of the intersection $R\cap R^\star$ is there a pair of matching EBC terms. Otherwise at least one (or both) end-point nodes sit outside $R$ or $R^\star$ respectively---in which case there is only one EBC term for the corresponding node set.

We can see that both types of summand are bounded above by unity. For in the second case, for any
$i,j\in R^\star$, 
\begin{eqnarray*}
\left|\frac{1\left[\left\{i,j\right\} \in E\right]}{T^{ij}}\right|
&\leq& 1\enspace,
\end{eqnarray*}
since 
$T^{ij}\geq 1$ since there is always a path through egonode $a$. The same holds for the case of $i,j\in R$ by definition. And in the first case of $i,j\in R\cap R^\star$, we have that
\begin{eqnarray*}
&& \left|\frac{1\left[\left\{i,j\right\} \in E\right]}{T^{ij}}-\frac{1\left[\left\{i,j\right\} \in E\right]}{T_1^{ij}}\right| \\
&\leq& 1 - \frac{1}{\max(\{|R^\star|,|R|\})-2} \\
&<& 1\enspace.
\end{eqnarray*}

Therefore it follows that
\begin{eqnarray*}
&& \left|\sum_{i,j \in R^\star} \frac{1\left[\left\{i,j\right\} \in E\right]}{T^{ij}}-\sum_{i,j \in R} \frac{1\left[\left\{i,j\right\} \in E\right]}{T_1^{ij}}\right| \\
&\leq& \left|\left\{i,j : i,j\in R\cap R^\star\right\}\right| \\
&& +\; \left|\left\{i,j : i\in R\backslash R^\star, j\in R \right\}\right| \\
&& +\; \left|\left\{i,j : i\in R^\star\backslash R, j\in R^\star \right\}\right| \\
&=& \mybinom {|R\cup R^\star|}{2}-|R^\star \backslash R||R\backslash R^\star| \\
&\leq& |R\cup R^\star|^2 / 2\\
&\leq& \left(|R\backslash R^\star|+|R^\star\backslash R|+|R^\star|\right)^2 / 2\\
&=& \left(|V| - q(R) + |R^\star|\right)^2 / 2\enspace,
\end{eqnarray*}
where the first equality follows by enumerating the $i,j$ pairs being counted as all those within $R$ or $R^\star$ provided that both nodes do not reside in separate set differences $R\backslash R^\star$ and $R^\star\backslash R$. 

This completes the result.

\section{Proof of Corollary\ref{cor:exp-mech-combined}}

The claim follows from Lemma~\ref{lem:mcSherry7} combined with Proposition~\ref{prop:disjoint-exp-utility}.

\section{Proof of Theorem~\ref{thm:utility-1}}\label{subsec:proofOfUtilityThm}

We begin by defining two events of interest on $R\in\Pow{V^-}$ for any $s>0$ to be chosen later: $S_s$ that our (centralised by Proposition~\ref{prop:disjoint-exp-utility}) exponential mechanism achieves near-optimality, and $T_s$ that the resulting EBC is near optimal.
Rewriting the event $S_s$ defined in Lemma 4, 
using $\tilde q(R)=|V|-q(R)$ and $\tilde q(R^\star)=0$, we have that 
$$S_s=\left\{R:\tilde q(R)\leq s\right\}\enspace,$$ 
which holds w.p. at least $1 - \exp(-\epsilon s / 2) 2^{-|V^-|}$. Next define 
$$T_s=\left\{R: |EBC-EBC_1|\leq \frac{1}{2}(s+|R^\star|)^2\right\}\enspace.$$
By Lemma 5 we have that
$|EBC-EBC_1|
\leq \frac{1}{2}(\tilde q(R)+|R^\star|)^2$
. And since $(a+b)^2/2$ is increasing in $a\in\mathbb{R}$ for $b\geq 0$, event $S_s$ implies then that 
$|EBC-EBC_1| \leq \frac{1}{2}(s+|R^\star|)^2 $ and so $S_s\subseteq T_s$ and $\Pr(T_s)\geq\Pr(S_s)$. 
Provided that $t> |R^\star|^2/2$, taking $s=\sqrt{2 t}-|R^\star|>0$ this completes the result.

\section{Analysis for Remark~\ref{remark:non-vacuous}}\label{subsec:nonVac}
 Consider a very sparse graph on nodes $|V|>10$, and consider a true ego network $R^\star$ with cardinality represented as a fraction $\alpha\in[0,1]$ of $|V|$ \ie $|R^\star|=\alpha |V|$. Let us now invoke the theorem with error bound $t$ taken to be a multiplier $\gamma>1$ of EBC---yielding a relative error bound of $|EBC-EBC_1|/EBC\leq \gamma$. 
For very sparse graphs, EBC can 
arbitrarily approach ${\mybinom{|R^\star|}{2}}\leq |R^\star|^2/2$ 
for $\gamma$ bounded away from zero and large $|V|$. Thus setting $t=\gamma|R^\star|^2/2$ corresponds to relative error $\gamma$ which is
meaningful for $\gamma$ not much larger than 1 (corresponding to private EBC within the same order of magnitude as non-private EBC). Next we wish to set the confidence $1-\delta$ to be very close to unity \eg $\delta=2^{-10}$ corresponds to confidence exceeding 99.9\%. With these choices, we set the confidence in Theorem 2 to $1-\delta$ and solve for the privacy budget required for \SubsetRelease:
$$ \epsilon \geq \frac{3\log_e 2}{(\gamma-1)\alpha}\enspace. $$

\section{Precise algorithms}
We rephrase the forward-pass algorithm from~\cite{roohi2019differentiallyprivate} as a generic subset release mechanism through 
Algorithms~\ref{alg:forward}, \ref{alg:itsampler} and~\ref{alg:pfsampler}.

\begin{algorithm}[ht]
 \caption{\SubsetRelease Two-Stage  \label{alg:forward} Sampler}
 \begin{algorithmic}[1]
 \renewcommand{\algorithmicrequire}{\textbf{Input:}}
 \REQUIRE public set $V$, private subset $R^\star\subseteq V$; $\epsilon>0$
  
  \STATE $I \longleftarrow \ITSampler(|V|, \epsilon)$
  \STATE $R \longleftarrow \PFSampler(V, R^\star, I)$
  \RETURN{ $R$} 
 
\end{algorithmic}
\end{algorithm}

\begin{algorithm}[H]
 \caption{\ITSampler \label{alg:itsampler}}
 \begin{algorithmic}[1]
 \renewcommand{\algorithmicrequire}{\textbf{Input:}}
 \renewcommand{\algorithmicrequire}{\textbf{Input:}}
 \REQUIRE cardinality $|V|$; $\epsilon>0$ \\
 \texttt{// Compute log-space PDF of $I$}
  \STATE $p_0 \longleftarrow -|V|\log\left(1 + \exp\left(\frac{\epsilon}{2}\right)\right)$
  \FOR{$i\in [|V|]$}
    \STATE $p_i \longleftarrow p_{i-1} + \log(|V|-i+1) - \log i$
  \ENDFOR \\
  \STATE $\psi \longleftarrow - \mathrm{Exp}(1)$
  \STATE $c \longleftarrow p_0$
  \FOR{$I\in [|V|]$}
    \IF{ $c\geq \psi$ }
      \RETURN{$I - 1$}
    \ENDIF
    \STATE $c \longleftarrow \log(\exp(c) + \exp(p_I))$
  \ENDFOR
  \RETURN{$I$}
\end{algorithmic}
\end{algorithm}
\begin{algorithm}[H]
 \caption{\PFSampler  \label{alg:pfsampler}}
 \begin{algorithmic}[1]
 \renewcommand{\algorithmicrequire}{\textbf{Input:}}
 \REQUIRE public set $V$; private subset $R^\star \subseteq V$; $I\in\{0,\ldots,|V|\}$ 
 \STATE $R \longleftarrow R^\star$
 \STATE $V_1,\ldots, V_{|V|-I} \sim \Unif{V}$ without replacement
 \FOR{$j\in [|V| - I]$}
   \IF{$V_j\in R$} 
     \STATE $R \longleftarrow R\backslash\{V_j\}$
   \ELSE
     \STATE $R \longleftarrow R\cup\{V_j\}$
   \ENDIF
 \ENDFOR
 \RETURN{$R$}

\end{algorithmic}
\end{algorithm}

\bibliographystyle{splncs04}
\bibliography{Bb1}

  \end{document}